# A Comparison of the Galactic Cosmic Ray Electron and Proton Intensities From ~1 MeV/nuc to 1 TeV/nuc Using Voyager and Higher Energy Magnetic Spectrometer Measurements – Are There Differences in the Source Spectra of These Particles?

**W.R. Webber and T.L. Villa**

New Mexico State University, Astronomy Department, Las Cruces, New Mexico, USA

## ABSTRACT

Using Leaky Box Model propagation calculations for H nuclei and a Monte Carlo diffusion propagation model for electrons, starting from specific source spectra, we have matched the observed LIS spectra of these cosmic rays measured by Voyager at lower energies and AMS-2 at higher energies, a range from ~10 MeV to ~1 TeV. The source spectra required are very similar rigidity spectra. Below ~6-10 GV the source spectra for both particles are $\sim P^{-2.25}$ and above 10 GV the spectra are $\sim P^{-2.36-2.40}$. This break in the source spectral index is not seen for He and C nuclei in a match of Voyager and AMS-2 intensities both of which have source rigidity spectra with an index ~-2.24 throughout the entire range of measured energies from ~10 MeV//nuc to ~1 TeV/nuc. The absolute source intensities of electrons and H nuclei are derived and the source ratio of accelerated electrons to H nuclei is between 2-5%. The total number of accelerated electrons is much greater than that for protons, however, because the accelerated electron spectrum extends down to ~1-2 MV rigidity whereas the H nuclei spectrum cannot be observed below ~50-100 MV because of ionization energy loss. Most of these low energy electrons escape from the galaxy forming an intergalactic background.

**Introduction**

In recent years there have been several magnetic spectrometer measurements of the cosmic ray electron and proton intensities from ~10 GeV/nuc to over 1 TeV/nuc near the Earth, above the energy at which solar modulation effects are believed to be a serious problem (BESS, Sanuki, et al., 2000; PAMELA, Adriani, et al., 2011; and AMS-2 Aguilar, et al., 2014, 2015a, 2015b). From these measurements the local interstellar (LIS) intensities can be reliably obtained with only a small correction. In spite of claims of accuracies of a few percent, however, these measurements have sometimes disagreed for both electron and proton intensities by as much as 10-20% or more leading to a resultant uncertainty in the actual electron and proton spectral indices above ~10 GeV.

The intensities of these particles have now also been measured on Voyager 1 in LIS space at energies $\geq 10^3$ times less than 10 GeV (Cummings, et al., 2016). Comparing these high and low energy intensities provides a large fulcrum on which to evaluate the uncertainties in the intensities and the spectra of these electrons and protons and other nuclei as well. In recent papers we have done just this type of comparison for H, He and C nuclei (Webber, 2017, 2018a).

For He and C nuclei we have found that the intensities measured at 10 MeV/nuc by Voyager and those at 1 TeV/nuc, measured by AMS-2, differing by a factor of $10^5$ in energy, are matched by a Leaky Box Model (LBM) diffusive propagation calculation to within a few percent at both the lowest and highest energies by assuming identical source rigidity spectra ~$P^{-2.24}$ from the lowest to the highest rigidity for these two nuclei, with a source ratio, He/C = 23.7, and propagation in the galaxy using a LBM with the mean path length $\lambda$=20.6 $\beta$ $P^{-0.45}$ above ~1.0 GV. Below 1.0 GV the mean path length was taken to be constant = 9 g/cm$^2$ (Webber, 2017).

Using the same propagation parameters noted above we have also matched the LIS H and He nuclei intensities between 10 MeV/nuc and 1 TeV/nuc (Webber, 2018a). If we use the AMS-2 measured H intensities above ~10 GeV and the V1 measured intensities below ~350 MeV, the low and high energy intensities can be matched to within a few percent, but this requires a distinct break in the H source energy/rigidity spectral index at between 5-10 GV, from one being ~$P^{-2.24}$ at lower rigidities, the same as that observed for He and C nuclei, to one ~$P^{-2.36}$, 0.12 power steeper, at higher energies/ rigidities.

However, if instead we use the PAMELA and BESS measurements of the higher energy H and He intensities, which differ from those of AMS-2 by 10-20%, we also can find a match between low and high energies/rigidities by now using the same H and He source spectra which are ~$P^{-2.24}$ throughout the entire rigidity range, the same as found earlier for He and C (Webber, 2018a). So in this case all 3 nuclei, H, He and C, would have the same ~$P^{-2.24}$ source spectrum at all rigidities.

In this paper we will make a comparison of the electron and H spectra, using the V1 electron data from ~1-60 MeV and the AMS-2 data on electrons >10 GeV and then comparing these new results for electrons with those for H nuclei, which were also made using the AMS-2 and Voyager data for each particle (Webber, 2018a).

## **The Data – The LIS Electron Spectrum**

In Figure 1 we show the LIS electron spectrum measured at V1 from ~1 MeV to ~60 MeV (Cummings, et al., 2016). The exponent of this LIS spectrum is ~$E^{-1.35}$ (or $P^{-1.35}$), where P = rigidity. These electrons appear to be part of a continuous spectrum extending up to ~1-10 GeV and beyond up to 1 TeV. To show the spectrum from 1 MeV to 10 GeV and above in Figure 1 we plot the electron intensities times $E^2$.

At the Earth this same electron spectrum is also measured from ~80 MeV to above 10 GeV by the PAMELA experiment (Adriani, et al., 2011, 2015). The results from this experiment are shown as blue points in Figure 1. At ~80 MeV, in between the maximum energy measured by Voyager and the minimum energy measured by PAMELA, (they do not quite overlap in energy) there is a factor of about 500 between the LIS intensity measured by V1 and the PAMELA intensity. This enormous solar modulation at the Earth is still a factor ~3-4 at 1 GeV as seen in Figure 1 and probably still 20-30% at 10 GeV, so it is important up to even 20 GeV, recognizing the claimed $\pm$ 1-2% accuracy of the experimental measurements above 10-20 GeV.

The spectral slope of the LIS electron spectrum much change drastically above a few hundred MeV (see Figure 1). Figure 2 shows the electron data at still higher energies from AMS-2, Aguilar, et al., 2014, with the electron intensities now multiplied by $E^3$. It is evident from this figure that the measured electron spectrum above 10-20 GeV is slightly steeper than $E^{-3.0}$, perhaps ~$E^{-3.1-3.2}$, up to about ~1 TeV so the spectral index must change from ~-1.35 below ~100 MeV to ~-3.1-3.2 above ~10 GeV.

In an earlier paper (Webber and Villa, 2017) we have shown that the measured V1 electron spectrum, which is ~$E^{-1.35}$ below ~60 MeV, is part of a continuous galactic electron spectrum extending up to ~10 GeV and above. This overall electron spectrum from a few MeV to 10 GeV can be well matched by assuming an electron source spectrum ~$E^{-2.25}$ (or $P^{-2.25}$) with essentially no change in the source exponent between the lower and higher energies up to ~10 GeV, along with an interstellar diffusion coefficient which is ~$P^{0.45}$ at rigidities above 1.0 GV changing to one that is ~$P^{-1.0}$ at lower rigidities. The $P^{0.45}$ dependence for the diffusion coefficient at higher rigidities is obtained from the AMS-2 measurement of the B/C ratio (Aguilar, et al., 2016) using a LBM calculation which fits the B/C ratio data from the AMS-2 experiment (Webber and Villa, 2016).

The observed changing LIS electron spectral index from -1.35 at low energies to ~3.1-3.2 at high energies can be explained as follows; synchrotron and inverse Compton energy losses, which are both ~$E^2$, steepen the source spectral index of -2.25 by ~0.8 to -3.05 above a few GeV. The change in the rigidity dependence of the diffusion coefficient from being ~ +0.45 above ~1.0 GV to being ~-1.0 below ~1.0 GV with the change occurring in the rigidity range from 0.2 to 0.5 GV, flattens the source exponent below ~0.5 GV by ~0.9 power in the exponent, from the exponent of the source spectrum which is ~-2.25, to a value of -1.35 (Webber and Villa, 2017). The appropriate galactic propagation calculations for electrons which lead to the above conclusions are made using a Monte Carlo diffusion model (Webber and Rockstroh, 1997; Webber and Higbie, 2008). The propagated electron spectra shown in Figures 1 and 2 will be discussed in the following section.

The remaining sections of this paper will describe this Monte Carlo calculation, the resultant LIS electron spectra and how they compare with the LIS H spectra, comparing them in terms of both energy and rigidity spectra (above a few GeV the two are equivalent since $\beta \rightarrow$ 1.0).

**The Monte Carlo Calculations of the LIS Electron Spectra**

The Monte Carlo diffusion model used here is described by Webber and Rockstroh, 1997, and has been used in subsequent publications (e.g., Webber, 2000; Webber and Higbie, 2008; Webber and Villa 2017). The key parameters for this calculation are; the thickness of the diffusion region, L = $\pm$1 Kpc, and the steps to the boundary = 4800 at 1.0GV. This leads to a diffusion coefficient, K = 1 x $10^{28}$ $cm^2 \cdot s^{-1}$ and a diffusion lifetime T = $L^2/2K$ = 2 x $10^7$ year. Another disk of 0.067 x L of average matter density = 1 particle/$cm^3$ is included. The source is uniform at Z = 0. All of the important loss terms are included; for synchrotron loss B = 6uG at Z = 0 and the inverse Compton term is between 0.5 and 0.7 times the synchrotron term depending on Z and energy.

In the example in Figure 1 we show the propagated LIS electron spectrum for a source electron spectrum ~$P^{-2.25}$ throughout the entire rigidity range from 1 MV to 10 GV and with an interstellar diffusion coefficient which is ~$P^{0.45}$ above 1.0 GV changing to ~$P^{-1.0}$ below 1.0 GV (see Webber and Villa, 2017). The source electron spectral index of -2.25 in this example matches the spectral index of -2.24 obtained earlier for the H source spectrum (and also the He and C source spectra) below 10 GV (Webber, 2018a). It is seen in Figure 1 that this propagated source electron spectrum provides an excellent match to the actual Voyager LIS measured electron spectrum between 1-60 MeV which has a slope = ~1.35 (e.g., ~$E^{-1.35}$), and at ~10 GeV it also provides a match which is about 20% higher than the AMS-2 data, a reasonable difference considering the fact that the AMS-2 measurement probably was made at a solar modulation level corresponding to ~20% solar modulation at 10 GeV (see Figure 2). This agreement between low and high energy electron intensities is achieved using a source spectrum which has an exponent independent of rigidity between 1 MeV and 10 GeV (no break) and with only one break in the rigidity dependence of the diffusion coefficient, at ~0.5 GV or less, a rigidity where, at the corresponding energies for nuclei ($\leq$ 20 MeV/nuc), it would be unobservable with present Voyager data. This is the scenario for electrons at energies up to ~10 GeV as described by Webber and Villa, 2017.

At energies >10 GeV the situation is best seen in Figure 2 where the intensities are x $E^3$. All measurements above 10 GeV (only the AMS-2 measurement is shown here) show an electron spectrum with an exponent >3.0 and a peak in the x $E^3$ spectrum at about 10 GeV.

It is also seen in Figure 2 that an electron source spectrum ~$P^{-2.25}$ which fits the observed LIS electron spectrum below 10 GeV, has a propagated LIS spectrum which, when extended above 10 GeV, has an exponent slightly less than 3.0 and therefore clearly no longer fits the AMS-2 data. The best fit for the measured AMS-2 high energy electron spectrum above 10 GeV requires a source spectrum with an exponent close to -2.40 above 10 GeV as shown in Figure 2. This would indicate a steepening by as much as 0.15 power in the source exponent at about 10 GeV. This source electron spectrum ~$P^{-2.40}$ fits the AMS-2 data above 10 GeV to within $\pm$10% at all energies up to the maximum ~500 GeV.

The key point here is that the source electron spectrum appears to change quite abruptly just below 10 GV from a slope ~$P^{-2.25}$ at lower rigidities to one ~$P^{-2.40}$ at higher rigidities. This change of slope is very similar to the change of slope that is necessary to fit the H spectrum using the AMS-2 measured H spectrum at high energies and the Voyager H data at low energies. From this low-high energy H nuclei comparison we have previously found that the H source spectrum also changes from a slope ~$P^{-2.24}$ at low rigidities, to a spectrum with a slope ~$P^{-2.36}$ at higher rigidities with the change of slope occurring at between 6-8 GV (Webber, 2018a).

Given the uncertainties in this comparison of electron and H nuclei source spectra including the rigidity of the spectral break which is $\pm$ 2 GV in each case and in which the solar modulation in the heliosphere plays an important role, we believe the source spectra of H nuclei and electrons could be essentially identical rigidity spectra, with a break in the spectral index of ~0.10-0.15, from -2.24 at low rigidities to -2.36-2.40 at higher rigidities, at a rigidity between 5-10 GV. Thus the spectra of these two charge one components, protons and electrons, would differ from those of the heavier cosmic ray nuclei with A/Z=2, of which He and C are examples, which are found to have source rigidity spectra ~$P^{-2.24}$ at all rigidities below 1 TeV, also based on comparisons of Voyager and AMS-2 data (Webber, 2017).

## **The Relative LIS Intensities of Electrons and H Nuclei**

In Figure 3 we show the LIS ratio of electron to H nuclei intensities in MeV, e/H (E), measured up to ~60 MeV at V1 outside the heliosphere (Cummings, et al., 2016) and also this ratio measured by AMS-2 above ~10 GeV where solar modulation is small (Aguilar, et al., 2014, 2015a). At 2 MeV this e/H (E) ratio is ~100 (yes 100!) decreasing to 1.10 at 60 MeV. This rapid decrease in the ratio is because of the increasing electron intensity at low energies where the electron spectrum is ~$E^{-1.35}$, compared to the decreasing H intensity which is caused mainly by ionization energy loss, which causes a peak in the differential H spectrum at ~30-40 MeV and then the decrease in H intensity at lower energies (see Figure 2 in Webber, 2018a).

At 10 GeV the AMS-2 direct measurement of the e/H(E) ratio is 0.012. This measured ratio decreases to ~0.001 at 1 TeV. Thus over the entire energy range from 2 MeV to nearly 1 TeV in energy the measured LIS e/H (E) ratio changes by a large factor ~$10^5$ from $10^2$ to $10^{-3}$. Yet we believe that the source e/H (P) ratio as a function of rigidity for these two particles should remain nearly constant since the source electron and H spectra appear to be nearly identical rigidity spectra with a possible break in the spectral index at a common rigidity between 5-10 GV.

This changing e/H (E) ratio as a function of energy can be explained as follows: We have already noted the relative effects of ionization energy loss on the H spectra below ~1 GeV. This along with the $1/\beta$ conversion between energy and rigidity which becomes important below ~1 GeV can explain most of the change in the e/H (E) ratio below about 1 GeV. Above 1 GeV the inverse Compton and synchrotron losses for electrons increase the electron spectral index by 0.8-0.9 from the source index of -2.25 to an observed index of ~3.1-3.2. This explains most of the changes in the e/H (E) ratio above a few GeV. There are no solar modulation free measurements of this ratio between 60 MeV ~10 GeV as noted earlier.

If we now compare the LIS Monte Carlo calculations of electron intensity shown in Figures 1 and 2 of this paper and listed in Table 1, with the LBM calculations of H nuclei intensities in Webber, 2018a, and also listed in Table 1, we obtain a calculated e/H (E) ratio which is also listed in Table 1 and is shown as a black line in Figure 3. This line fits remarkably well the observed e/H (E) ratio measured by Voyager at low energies and by AMS-2 at high energies. At 1 GeV this e/H (E) ratio is 0.058 and the absolute values of the LIS intensities are 0.014 for electrons and 2.4 for protons in units of $P/m^2 \cdot sec \cdot$ MeV.

The equivalent LIS rigidity spectra for e and H nuclei are shown in Figure 4. Note the large excess of LIS electron intensities over those for hydrogen at low rigidities.

## The Relative Source (Accelerated) Spectra and Intensities of e and H Nuclei and the e/H (P) Ratio

From the respective calculations of the LIS e and H nuclei spectra described above we recognize that the source spectra of these two components are ~$P^{-2.24-2.25}$ below 6-10 GV, with the exponent in each case increasing to $P^{-2.36-2.40}$ above 10 GV. The absolute values of the constants that determine these two source spectra depend on the loss of particles from the assumed galactic volume. These losses include nuclear interactions for H, escape from the diffusing region for both e and H, ionization E loss also for both e and H, and bremsstrahlung loss plus Synchrotron and Inverse Compton loss for electrons. In the Monte Carlo diffusion model all of these losses are tabulated so that we know the fractions of injected particles remaining in the galaxy (diffusing volume) for each energy.

These e/H fractions of particles remaining in the galaxy may be used to determine the absolute source intensities. For example, at 1 GV this surviving fraction is 71% for electrons and 55% for H nuclei. These estimated source spectra for electrons and H nuclei, which are based on the surviving fractions as a function of rigidity, are shown in Figure 5. The e/H(P) ratio for these spectra is within the range 2-5%.

## Summary and Conclusions

In this paper we have used a Monte Carlo diffusion model to determine the electron source spectrum that will, after propagation, provide a match for the observed LIS electron spectrum between 1-60 MeV measured at Voyager and above 10 GeV measured at AMS-2 near the Earth. This electron source spectrum is a rigidity spectrum ~$P^{-2.24-2.25}$ below ~6-10 GV and ~$P^{-2.36-2.40}$ above ~10 GV. A key feature of this analysis is the $E^{-1.35}$ spectrum that is measured at Voyager at low energies. This low energy electron spectrum is well matched starting with a source spectrum ~$P^{-2.25}$ along with diffusion loss from the diffusion volume of the galaxy which is ~$P^{-1.0}$ and which flattens the source spectral index by about 0.9 power, thus producing the spectrum observed at Voyager. This electron source spectrum with a break ~0.12-0.15 at between 6-10 GV is very similar to the source spectrum we have recently derived for H nuclei using AMS-2 data (Webber, 2018a).

The relative amplitude of these two spectra is such that the e/H(P) ratio at the source is within the range 2-5%. This is consistent with the possible ionization state of the medium in which the cosmic rays are accelerated.

In our efforts to compare Voyager measurements at low energies with those from AMS-2 above 10 GeV, we have, from the observed LIS spectra over a wide energy range from 10 MeV to 1 TeV, now determined the source spectra for four cosmic ray particles; electrons, H, He and C nuclei (Webber, 2017, Webber 2018a and this paper). For He and C these source spectra appear to be rigidity spectra and to have the same spectral index = -2.24 $\pm$ 0.02 between the lowest energies of a few MeV/nuc up to almost 1 TeV/nuc. For both electrons and H nuclei, however, the source spectral index appears to be the same $P^{-2.25}$ dependence below 6-10 GV, but steepens by 0.10-0.15 power in the index at higher rigidities.

**Acknowledgements:** The authors are grateful to the Voyager team that designed and built the CRS experiment with the hope that one day it would measure the galactic spectra of nuclei and electrons. This includes the present team with Ed Stone as PI, Alan Cummings, Nand Lal and Bryant Heikkila, and to others who are no longer members of the team, F.B. McDonald and R.E. Vogt. Their prescience will not be forgotten. This work has been supported throughout the more than 40 years since the launch of the Voyagers by the JPL.

| TABLE 1<br>LIS Intensities of Electrons and H Nuclei | | | |
|---|---|---|---|
| **Energy (MeV)** | **H1*** | **Electrons*** | **etrons/H1 Ratio** |
| 1.76 | 1.14E+01 | 2.00E+03 | 1.75E+02 |
| 2.37 | 1.40E+01 | 1.63E+03 | 1.16E+02 |
| 3.16 | 1.68E+01 | 1.07E+03 | 6.37E+01 |
| 4.27 | 1.97E+01 | 7.32E+02 | 3.72E+01 |
| 5.6 | 2.29E+01 | 4.66E+02 | 2.03E+01 |
| 7.5 | 2.60E+01 | 3.22E+02 | 1.24E+01 |
| 10 | 2.84E+01 | 2.19E+02 | 7.71E+00 |
| 13 | 3.09E+01 | 1.51E+02 | 4.89E+00 |
| 18 | 3.26E+01 | 1.03E+02 | 3.16E+00 |
| 24 | 3.37E+01 | 7.06E+01 | 2.09E+00 |
| 32 | 3.40E+01 | 4.91E+01 | 1.44E+00 |
| 42 | 3.25E+01 | 3.29E+01 | 1.01E+00 |
| 56 | 3.05E+01 | 2.22E+01 | 7.28E-01 |
| 75 | 2.78E+01 | 1.48E+01 | 5.32E-01 |
| 100 | 2.39E+01 | 1.00E+01 | 4.18E-01 |
| 133 | 2.02E+01 | 6.64E+00 | 3.29E-01 |
| 178 | 1.59E+01 | 4.33E+00 | 2.72E-01 |
| 237 | 1.22E+01 | 2.80E+00 | 2.30E-01 |
| 316 | 9.42E+00 | 1.79E+00 | 1.90E-01 |
| 422 | 7.25E+00 | 1.00E+00 | 1.38E-01 |
| 562 | 5.24E+00 | 5.40E-01 | 1.03E-01 |
| 750 | 3.42E+00 | 2.85E-01 | 8.33E-02 |
| 1000 | 2.14E+00 | 1.43E-01 | 6.68E-02 |
| 1334 | 1.26E+00 | 6.34E-02 | 5.03E-02 |
| 1779 | 7.89E-01 | 3.32E-02 | 4.21E-02 |
| 2372 | 4.47E-01 | 1.53E-02 | 3.42E-02 |
| 3163 | 2.53E-01 | 7.00E-03 | 2.77E-02 |
| 4218 | 1.39E-01 | 3.12E-03 | 2.24E-02 |
| 5624 | 7.44E-02 | 1.42E-03 | 1.91E-02 |
| 7500 | 3.95E-02 | 6.63E-04 | 1.68E-02 |
| 10001 | 2.01E-02 | 2.49E-04 | 1.24E-02 |
| 13337 | 9.98E-03 | 1.05E-04 | 1.05E-02 |
| 17785 | 4.82E-03 | 4.12E-05 | 8.55E-03 |
| 23717 | 2.32E-03 | 1.61E-05 | 6.94E-03 |
| 31627 | 1.08E-03 | 6.57E-06 | 6.08E-03 |
| 42176 | 4.95E-04 | 2.60E-06 | 5.25E-03 |
| 56242 | 2.27E-04 | 1.08E-06 | 4.76E-03 |
| 75000 | 1.03E-04 | 4.20E-07 | 4.08E-03 |
| 100014 | 4.63E-05 | 1.64E-07 | 3.54E-03 |
| 133371 | 2.09E-05 | 6.54E-08 | 3.13E-03 |
| 177853 | 9.44E-06 | 2.54E-08 | 2.69E-03 |
| 237171 | 4.23E-06 | 1.07E-08 | 2.53E-03 |
| 316273 | 1.81E-06 | 4.10E-09 | 2.27E-03 |
| 421757 | 8.14E-07 | 1.53E-09 | 1.88E-03 |
| 562422 | 3.62E-07 | 6.06E-10 | 1.67E-03 |
| 750001 | 1.62E-07 | 1.38E-10 | 1.23E-03 |
| 1000143 | 7.27E-08 | 8.92E-11 | 8.52E-04 |

*Intensities are in P/m$^2$·sr·s·MeV

## FIGURE CAPTIONS

**Figure 1:** Measurements of the low energy LIS galactic cosmic ray electron spectrum at Voyager (Cummings, et al., 2016). The high energy electron spectra measured near the Earth by PAMELA (Adriani, et al., 2015) and by AMS-2 (Aguilar, et al., 2014) are also shown. A calculated LIS spectrum using a Monte Carlo program for galactic propagation with a source electron spectrum ~$P^{-2.25}$ is also shown.

**Figure 2:** A comparison of the measured AMS-2 high energy electron spectrum and that calculated using a Monte Carlo galactic diffusion model with an electron source spectrum ~$P^{-2.25}$ below 7.5 GV and a source spectral index that changes to $P^{-2.40}$ above P=7.5 GV. The shaded region at lower rigidities gives an idea of the effect of solar modulation on the electron spectrum at the time of the AMS-2 measurement.

**Figure 3:** Measurements (in red) and propagation calculations of the LIS e/H (E) intensity ratio as a function of energy. The V1 LIS measurements are shown from ~2 to 60 MeV, where the ratio changes from ~100 to ~1.0. The AMS-2 measurements are from 10 GeV to 1 TeV where the ratio changes from 0.012 to 0.001. The calculated LIS ratio at 1 GeV is based on the LIS electron spectrum derived in this paper and the LIS proton spectrum derived in Webber, 2018a (see Table I).

**Figure 4:** The calculated LIS rigidity spectra for electrons and H nuclei that match the observed Voyager intensities at low rigidities and the AMS-2 intensities at higher rigidities. The calculations are described in the text.

**Figure 5:** The source spectra required to reproduce the observed LIS intensities of electrons and H nuclei measured by Voyager and by AMS-2 as shown in earlier figures. Note that at 1 GV this source e/H ratio is 0.02-0.03.

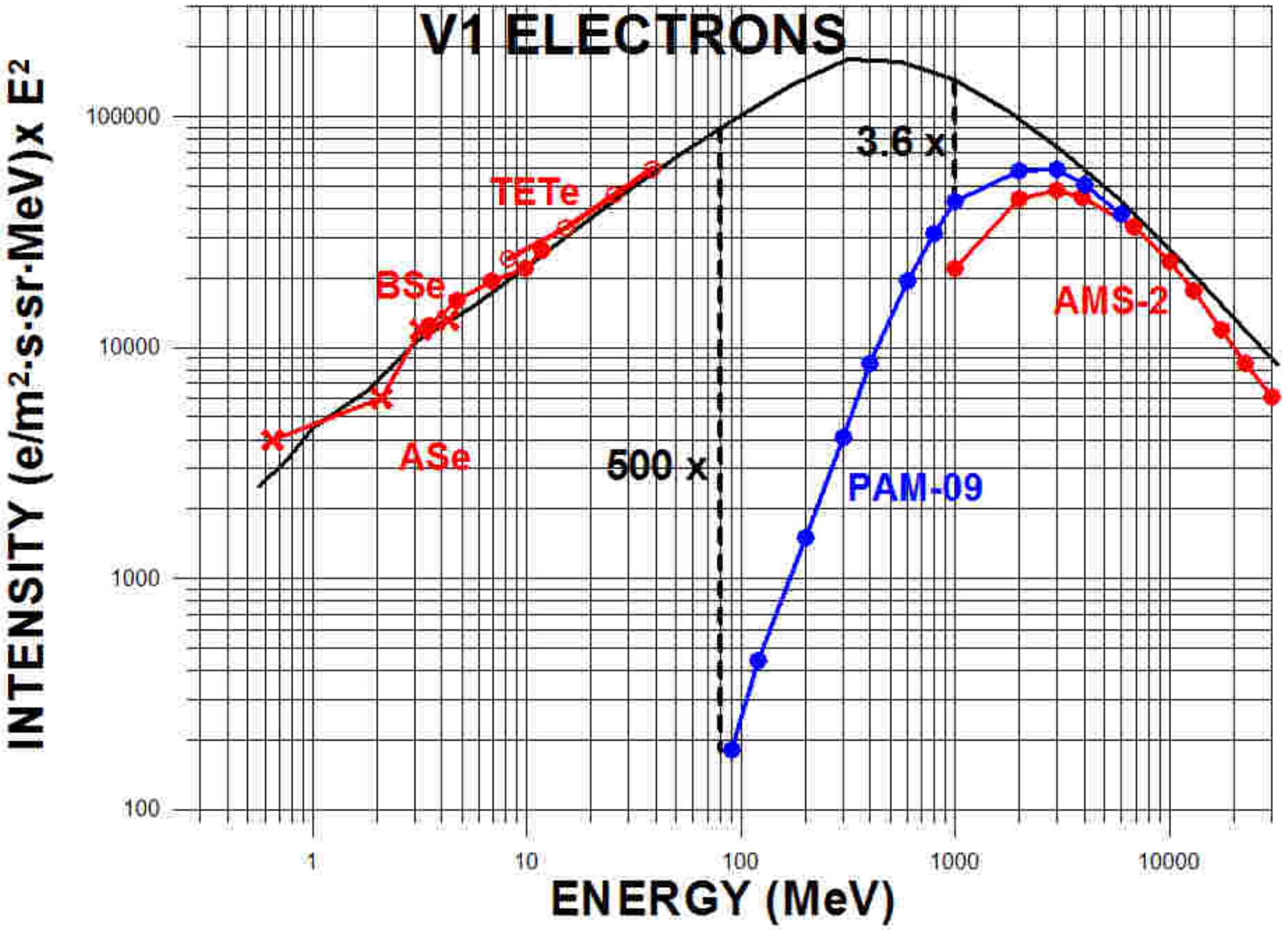

FIGURE 1

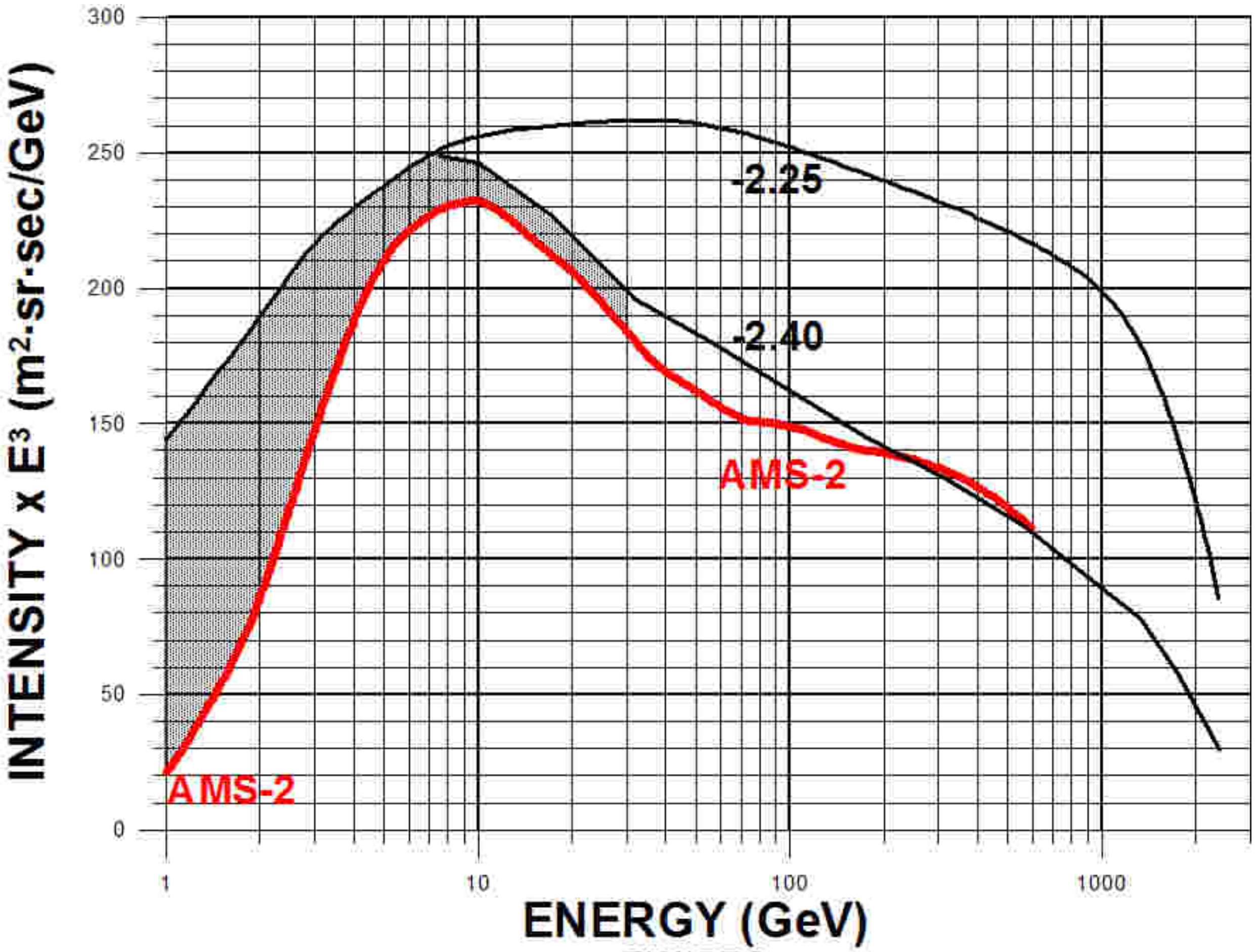

FIGURE 2

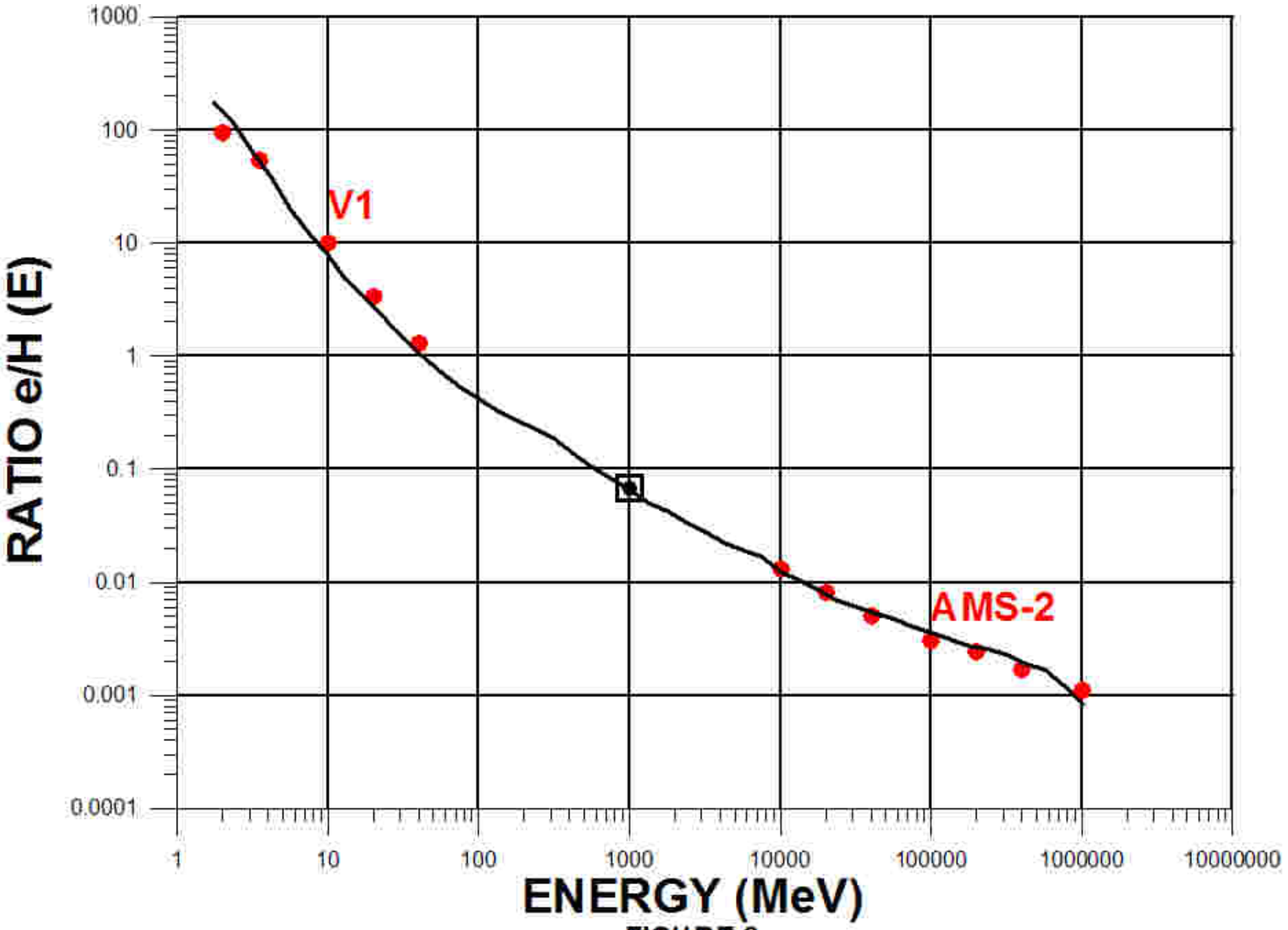

FIGURE 3

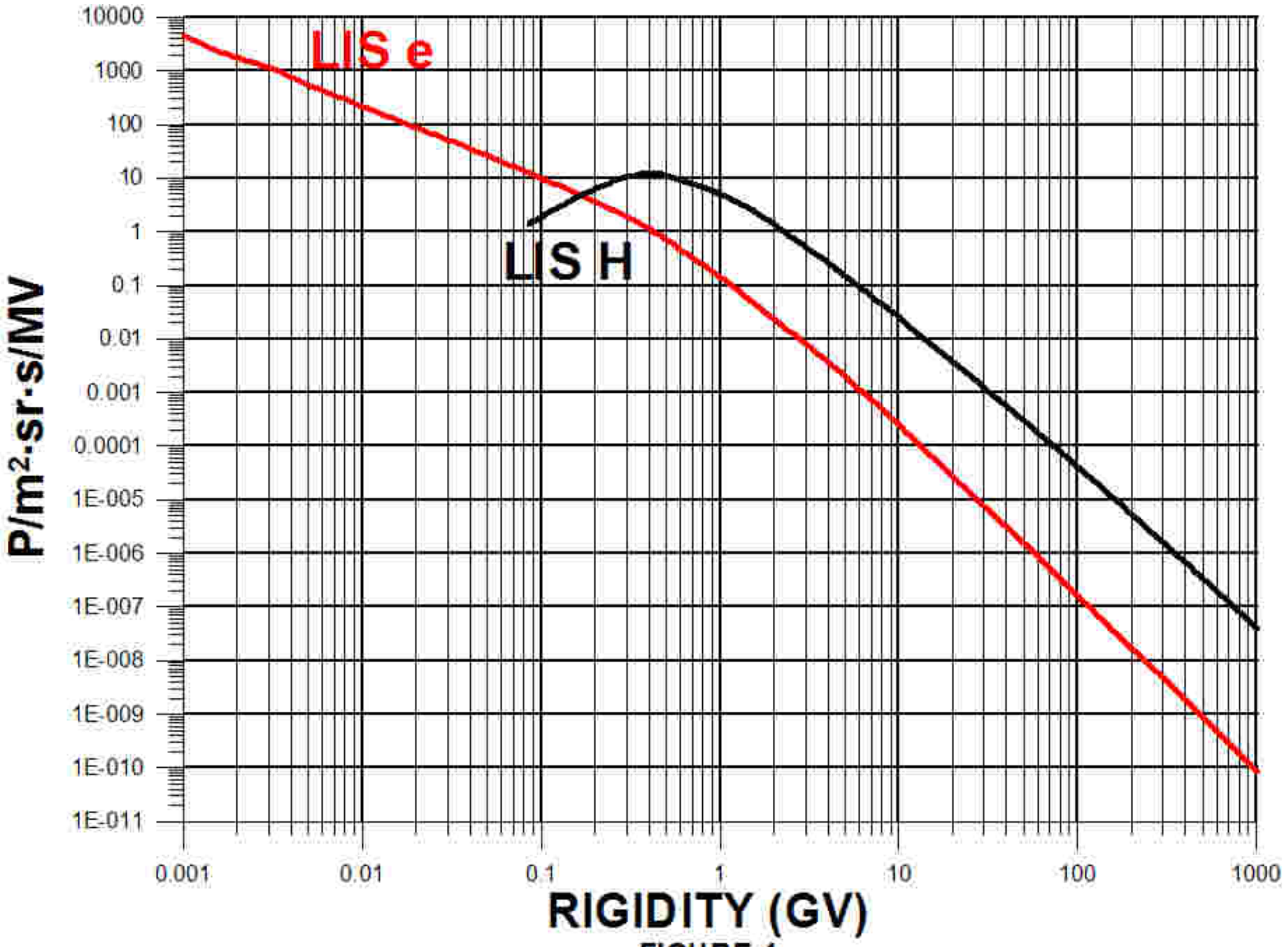

FIGURE 4

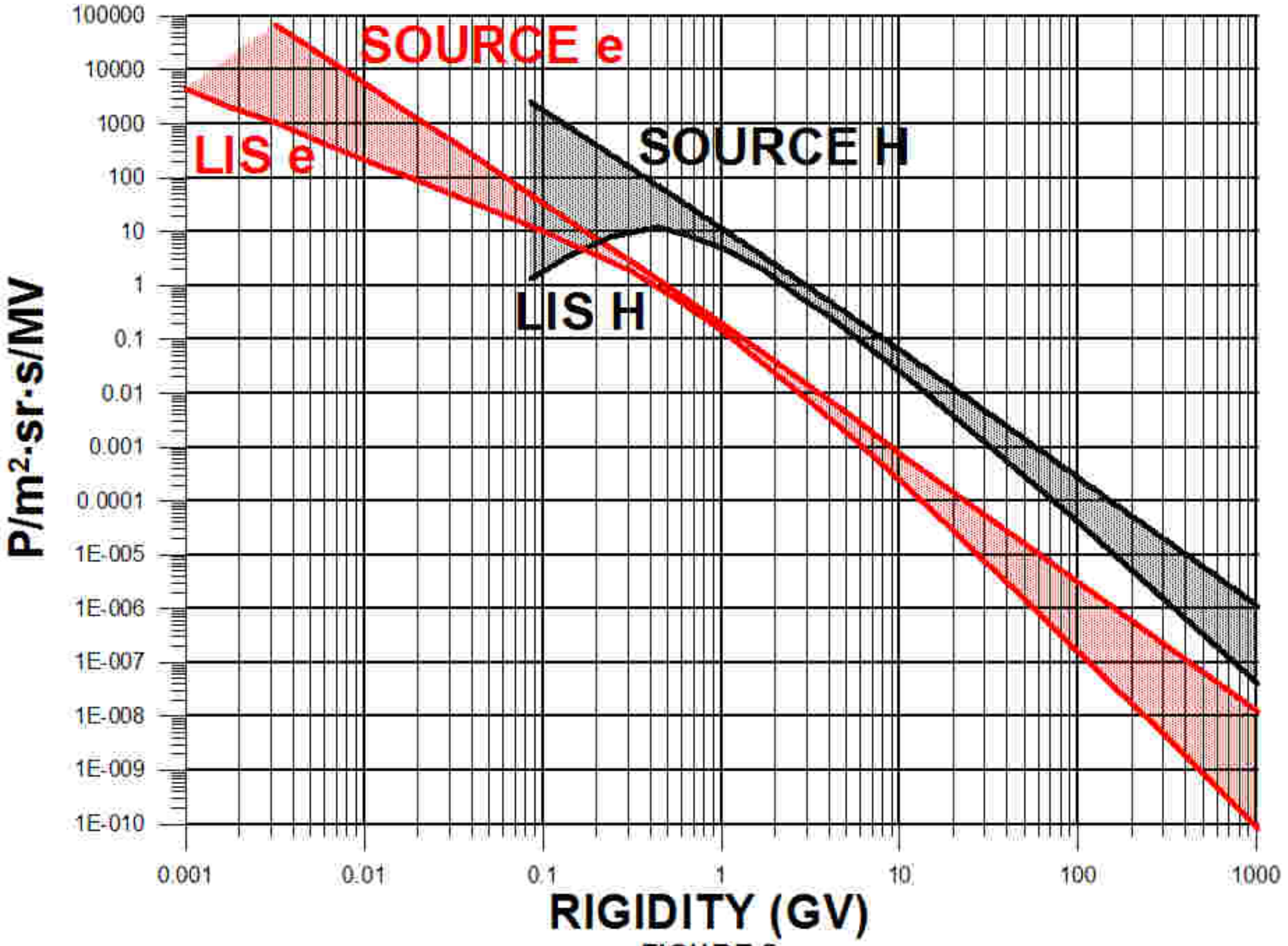

FIGURE 5